\documentclass{PoS}

\usepackage{amsfonts}
\usepackage{amsmath}
\usepackage{amssymb}
\usepackage{amsthm}
\usepackage{array}
\usepackage{dcolumn}
\usepackage{multirow}
\usepackage{verbatim}
\usepackage{graphicx}
\usepackage{color}
\usepackage{stackrel}

\title{The gradient flow in simple field theories}

\ShortTitle{The gradient flow in simple field theories}

\author{\speaker{Christopher Monahan}\thanks{Current address: New High Energy 
Theory Center, Rutgers, The State University of New Jersey, 136 Frelinghuysen 
Road, Piscataway, NJ 08854-8019}\\
Department of Physics and Astronomy, University of Utah,
Salt Lake City, Utah 84112, USA\\
        E-mail: \email{chris.monahan@rutgers.edu}}

\abstract{The gradient flow is a valuable tool for the lattice 
community, with applications from scale-setting to implementing chiral 
fermions. Here I focus on the 
gradient flow as a means to suppress power-divergent mixing. Power-divergent 
mixing stems from the hypercubic symmetry of the lattice regulator and is a 
particular difficulty for calculations of, for example, high moments of parton 
distribution functions. 
The gradient flow 
removes power-divergent mixing on the lattice, provided the flow time is kept 
fixed in physical units, at the expense of introducing a new physical 
scale in the continuum. One approach to dealing with this new scale is the
smeared operator product expansion, a formalism 
that systematically connects 
nonperturbative calculations of flowed operators to continuum physics.
I study the role of the gradient flow in suppressing power-divergent 
mixing and present the first nonperturbative study in scalar field 
theory.}

\FullConference{The 33rd International Symposium on Lattice Field Theory\\
		14 -18 July 2015\\
		Kobe International Conference Center, Kobe, Japan*}

\begin{document}

\section{Introduction}

The curse of power-divergent mixing afflicts a range of lattice calculations, 
but the prototypical example occurs in 
calculations of matrix elements of twist-2 operators, 
where twist is the dimension minus the spin of the operator. These matrix 
elements arise, for example, in lattice determinations of moments of 
parton distribution functions, which capture the distribution of a fast-moving 
nucleon's longitudinal momentum amongst its constituents. For twist-2 
operators, power-divergent mixing stems 
from the hypercubic symmetry of the lattice regulator.

In general, the symmetry restrictions of the finite hypercubic 
group are less stringent than those imposed by the orthogonal group of 
continuous Euclidean spacetime and allow 
for more complicated radiative mixing. Operators of 
definite angular momentum, for example, do not mix in the continuum. The 
lattice regulator breaks rotational symmetry, however, and angular 
momentum is no longer a good quantum number in a discretised spacetime. Lattice 
operators that correspond 
to a state with definite angular momentum in the continuum may therefore mix. 
For operators of different mass dimension, this mixing generates contributions 
that diverge as an inverse power of the lattice spacing in the continuum 
limit: the problem of power-divergent mixing. Such contributions must be 
removed to extract meaningful continuum physics.

In \cite{Monahan:2015lha}, we introduced a modified operator product 
expansion--the 
\emph{smeared operator product expansion}--to remove 
power-divergent mixing via the gradient flow. The gradient flow, a 
gauge-invariant form of smearing with particularly useful renormalisation 
properties \cite{Luscher:2011bx}, drives the 
original degrees of freedom to the stationary 
points of the action and corresponds to a continuous stout-smearing procedure 
\cite{Morningstar:2003gk}. 
Smearing has long been used in lattice calculations 
to reduce discretisation effects and improve the continuum limit of lattice 
data. There is a history, too, of modifying the 
operator product expansion on 
the lattice to account for power-divergent mixing 
\cite{Dawson:1997ic}, although these techniques have not been widely adopted. 
Here
I discuss the use of the gradient flow to remove power-divergent mixing and 
consider some examples in scalar field theory.

\section{The smeared operator product expansion}
The operator product expansion (OPE) for a non-local operator in scalar field 
theory is 
widely-known (see, for example, \cite{Collins:1984rdg}), so here I just 
introduce some necessary notation. I write the OPE for a non-local 
operator, ${\cal Q}(x)$, as
\begin{equation}
{\cal Q}(x)  \stackrel{x\rightarrow 0}{\sim} \; \sum_k 
c_k(x,\mu) {\cal O}_R^{(k)}(0,\mu).
\end{equation}
The $c_k(x,\mu)$ are perturbative Wilson coefficients that 
capture the short-distance physics associated with the renormalised
local operator ${\cal O}_R^{(k)}(0,\mu)$, where $\mu$ is the renormalisation 
scale. This operator is a polynomial in the scalar 
field and its derivatives, and the operator's free-field mass dimension governs 
the leading spacetime dependence of the Wilson coefficients. Radiative 
corrections generate sub-leading dependence on the spacetime separation. The 
Wilson coefficients are functions of the spacetime separation $x$; the 
(renormalised) mass $m$; and the renormalisation scale, $\mu$. Here, one should 
interpret this equality in the weak sense of holding between matrix elements.

In \cite{Monahan:2015lha} we proposed a new expansion in terms 
of smeared operators, the smeared OPE:
\begin{equation}
{\cal Q}(x)  \stackrel{x\rightarrow 0}{\sim} \; \sum_k 
d_k(\tau,x,\mu) {\cal S}_R^{(k)}(\tau,0,\mu).
\end{equation}
The smeared coefficients $d_k(\tau,x,\mu)$ are now functions of three 
scales: the smearing scale, 
$\tau$; the spacetime separation, $x$; and the 
renormalisation scale, $\mu$. The smeared operator, ${\cal 
S}_R(\tau,0)$, has the same free-field mass dimension as its local 
counterpart. Thus, the smeared coefficients exhibit the same 
leading spacetime dependence as that of the local Wilson coefficients.

Although we referred to these operators as generically ``smeared'', one 
should recognise that we had in mind a particular form of smearing: the 
gradient flow \cite{Luscher:2011bx,Narayanan:2006rf}. The 
gradient flow is a classical evolution of the original degrees of freedom 
towards the stationary points of the action in a new dimension, the flow time, 
with the particularly useful property that renormalised correlation functions 
remain renormalised at non-zero flow time (up to a fermion 
renormalisation) \cite{Luscher:2011bx}.

Working with scalar field theory in two dimensions, defined by the action
\begin{equation}
S_\phi[\phi] = \frac{1}{2}\int \mathrm{d}^2x\, \left[(\partial_\nu  
\phi)^2 + m_\phi^2\phi^2 + \frac{\lambda_\phi}{2}\phi^4\right],
\end{equation}
the flow time evolution takes a particularly simple form (see 
\cite{DallaBrida:2015tty} for a rather different application of the gradient 
flow that also takes advantage of its simplicity):
\begin{equation}\label{eq:flowevo}
\frac{\partial \rho(\tau,x)}{\partial \tau}= \partial^2 \rho(\tau,x).
\end{equation}
Here $\partial^2$ is the Euclidean Laplacian operator. Imposing the Dirichlet 
boundary condition $\rho(0,x) = \phi(x)$, one can solve this equation exactly. 
The solution is
\begin{equation}
\rho(\tau,x) = \frac{1}{4\pi\tau}
\int\mathrm{d}^2y\,e^{-(x-y)^2/4\tau}\phi_2(y),
\end{equation}
which demonstrates explicitly the ``smearing'' effect of the gradient flow: the 
flow time exponentially damps ultraviolet fluctuations.  The root-mean-square 
smearing length, which is $s_{\mathrm{rms}} = 2\sqrt{\tau}$ in two dimensions, 
characterises the corresponding smearing radius.

The flow evolution equation, Equation \eqref{eq:flowevo}, is 
straightforward and manifestly corresponds to Gaussian-smearing of the scalar 
degrees of freedom. In principle, however, we are free to choose the precise 
form of the flow time evolution equation (provided it drives the fields to 
configurations corresponding to minima of the action), and we could incorporate 
interactions in the flow time evolution, as is done in QCD and the nonlinear 
sigma model. This choice would, unfortunately, remove the renormalisation 
property of the gradient flow that proves so useful: renormalised correlation 
functions would no longer be guaranteed to remain finite at non-zero flow time. 
Ultimately this result stems from the internal symmetries of QCD and the 
nonlinear sigma model that have no analogue in scalar $\phi^4$ theory\footnote{I 
am indebted to M.~Dalla Brida and R.~Brower for discussions regarding this 
point.}. In pure Yang-Mills theory and the nonlinear sigma model, it is gauge 
invariance, manifest through appropriate BRST symmetries, that ensures no new 
counterterms can be generated by the gradient flow 
\cite{Luscher:2011bx}. 

\section{Nonperturbative scalar field theory}

I study two-dimensional $\phi^4$ theory on small ($N = L/a = 16, 32$) to 
medium-sized square lattices ($N = 128, 256$). 
The lattice action is
\begin{equation}
S_\phi^{\mathrm{latt}}[\phi] = \sum_{n}\left[- \sum_{i=1}^2 \phi(n+i)\phi(n) + 
\left(2+\frac{m_0^2}{2}\right) \phi^2(n)  + \frac{\lambda}{4} \phi^4(n)\right],
\end{equation}
where the sum runs over all lattice sites $n$ in the lattice 
volume $N^2 = (L/a)^2$ and $m_0 = am_\phi$ is the dimensionless bare mass. The 
lattice coupling constant $\lambda = a^2\lambda_\phi$ is also dimensionless.
In the infinite volume limit, 
this theory has a continuous phase transition between the symmetric phase, in 
which 
$\langle \phi \rangle = 0$, and a broken phase, in which $\langle \phi \rangle 
\neq 0$, explored in, for example, 
\cite{Loinaz:1997az}. The continuum limit corresponds to the origin in the 
bare lattice mass-coupling plane and is parameterised by a single 
dimensionless number.

In two dimensions, $\phi^4$ theory is super-renormalisable \cite{Loinaz:1997az}; 
there is a 
single divergent tadpole diagram, which is given by
\begin{equation}
A_{m_0^2}^{\mathrm{latt}} = \frac{1}{N^2}\sum_{k=0}^{N -1}
\frac{1}{4\sin^2(\pi 
k/N)+m^2} \;\stackrel[m\,\mathrm{fixed}]{N\rightarrow\infty}{=} \;
\int_0^\infty\mathrm{d} u\, e^{-u(4- m^2)} 
[I_0(2u)]^2, 
\end{equation}
where the second equality holds in the infinite volume limit, $I_0(z)$ is the 
modified Bessel function of the first kind, and $m$ is the 
renormalised (lattice) mass. Perturbatively, the renormalisation condition 
$\delta m^2 
= 3\lambda A_{m^2}$ removes all ultraviolet divergences. In principle one could 
choose to define a renormalised coupling, either through 
the four-point function or with a finite volume scheme based on the 
gradient flow, analogous to that proposed in \cite{Fodor:2012td}, but this is 
unnecessary, because the 
renormalised coupling differs by a finite shift from the bare 
coupling.

\paragraph{Numerical tests}

To combat critical slowing down, I use a Monte Carlo procedure that 
incorporated five Metropolis update steps followed by an embedded Wolff 
cluster update step \cite{Wolff:1988uh} (there also exist microcanonical and 
multigrid methods for this system \cite{Morningstar:2007zm}). I thermalised the 
lattices for at least $10^4$ such ``sweeps'' and took measurements every 100 
sweeps to reduce autocorrelations (at the critical mass, for example, I found 
integrated autocorrelation times up to approximately 80 update steps for 
$\langle \phi^2 \rangle$, in line with \cite{Loinaz:1997az}). 

\begin{figure}
\centering
\hspace*{-5pt}\begin{minipage}{0.258\textwidth}
\includegraphics[width=\textwidth,height=0.16\textheight]
{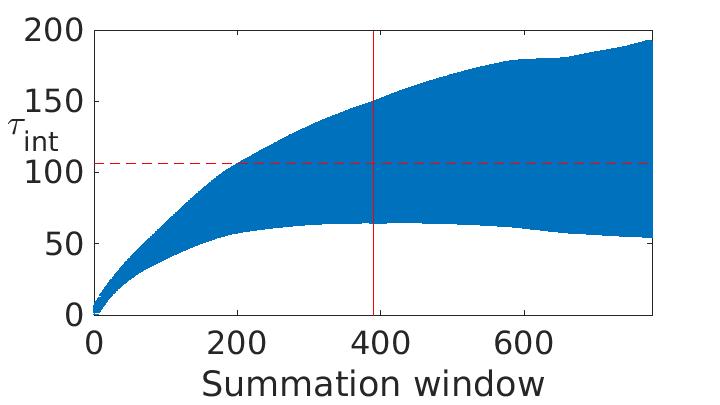}
\end{minipage}%
\begin{minipage}{0.258\textwidth}
\includegraphics[width=\textwidth,height=0.16\textheight]
{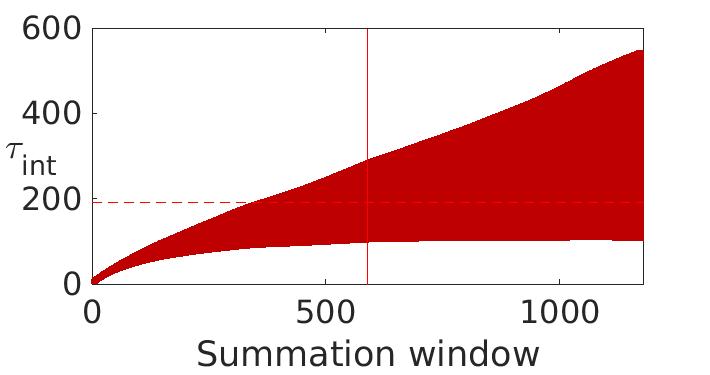}
\end{minipage}%
\begin{minipage}{0.258\textwidth}
\includegraphics[width=\textwidth,height=0.16\textheight]
{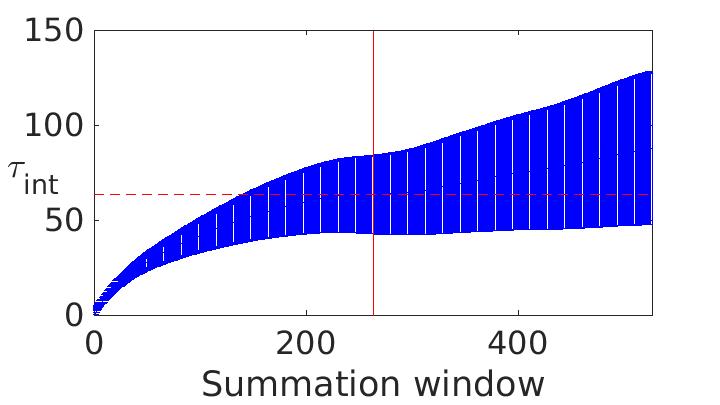}
\end{minipage}%
\begin{minipage}{0.258\textwidth}
\includegraphics[width=\textwidth,height=0.16\textheight]
{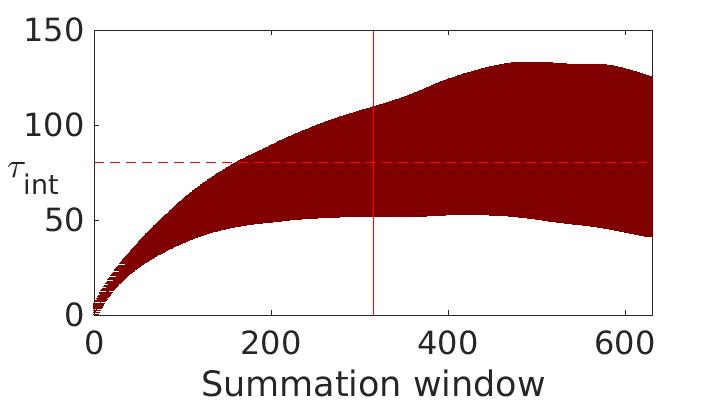}
\end{minipage}%
\caption{\label{fig:tauint} 
The autocorrelation time for $\langle \phi^2 \rangle$, as a function of the 
summation window of the Gamma analysis of \cite{Wolff:2003sm}. 
Simple Metropolis updating at: ({\em far left}) $m_0^2 = -0.68$ and ({\em 
centre left}) 
 $m_0^2 = -0.72$.   
Metropolis algorithm with 
an embedded cluster update step at: ({\em centre right}) $m_0^2 = -0.68$ and 
({\em far right}) $m_0^2 = -0.72$. All plots show data for $N = 64$ and 
$\lambda 
= 0.5$. The vertical red 
lines represent the value of the summation 
window automatically chosen by the Gamma analysis procedure and the horizontal 
red lines give the corresponding integrated autocorrelation time. }
\end{figure}

I plot the 
integrated autocorrelation 
times $\tau_{\mathrm{int}}$, calculated using the Gamma analysis of 
\cite{Wolff:2003sm}, for $\langle \phi^2 \rangle$ with simple Metropolis 
updating in the left-hand panels of Figure \ref{fig:tauint} and for Metropolis 
steps combined with a embedded cluster update step in the right-hand panels of 
the same figure (for details, see the corresponding caption). In general, 
including
the cluster update improves the integrated autocorrelation time by a factor of 
two to three. The 
increased integrated autocorrelation time at $m_0^2 = -0.72$ indicates the 
critical slowing down associated with a phase transition.

\begin{figure}
\hspace*{-5pt}\begin{minipage}{0.258\textwidth}
\includegraphics[width=\textwidth,height=0.16\textheight]
{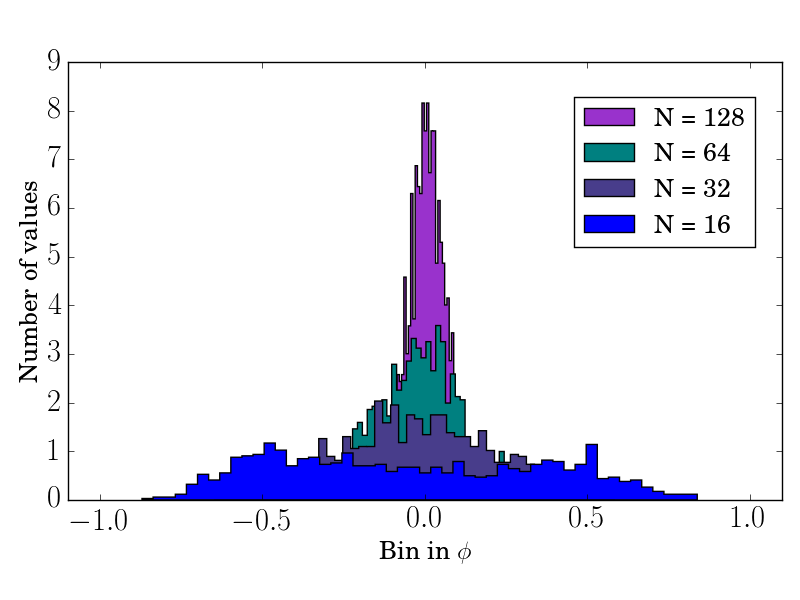}
\end{minipage}%
\begin{minipage}{0.258\textwidth}
\includegraphics[width=\textwidth,height=0.16\textheight]
{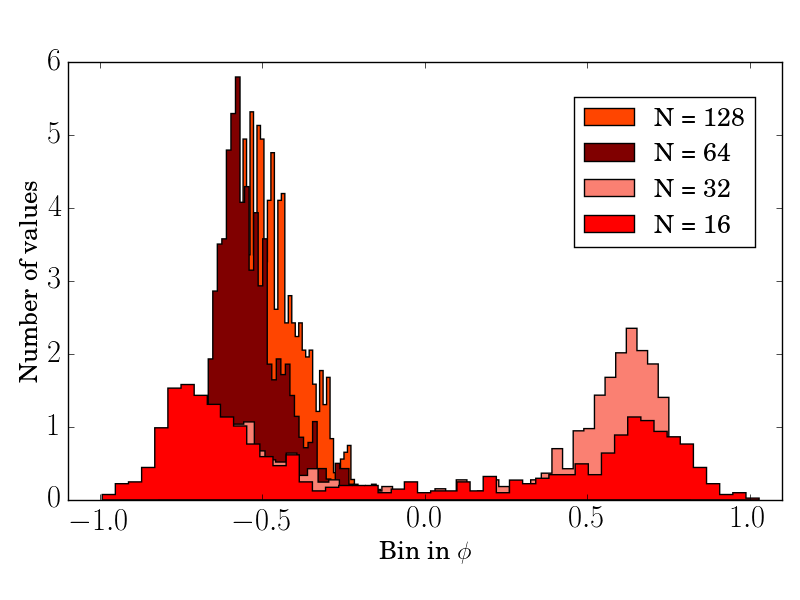}
\end{minipage}%
\begin{minipage}{0.258\textwidth}
\includegraphics[width=\textwidth,height=0.16\textheight]
{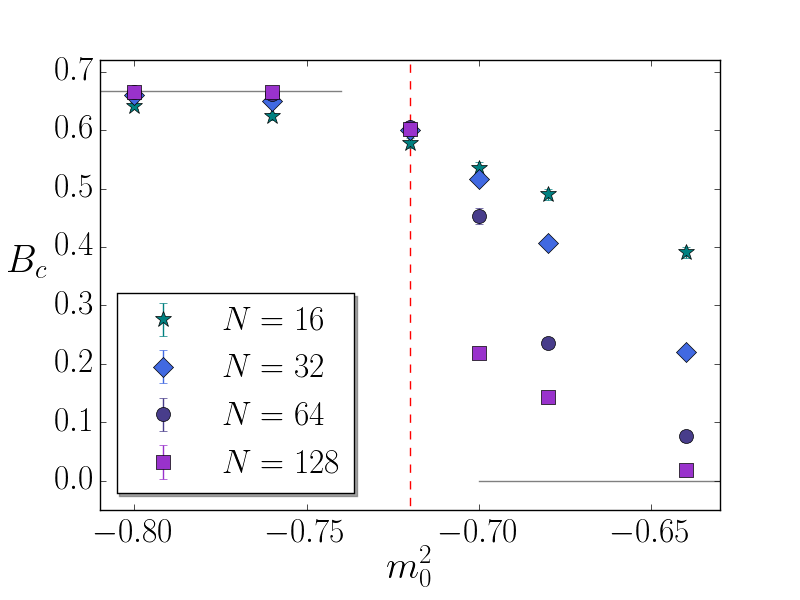}
\end{minipage}%
\begin{minipage}{0.258\textwidth}
\includegraphics[width=\textwidth,height=0.16\textheight]
{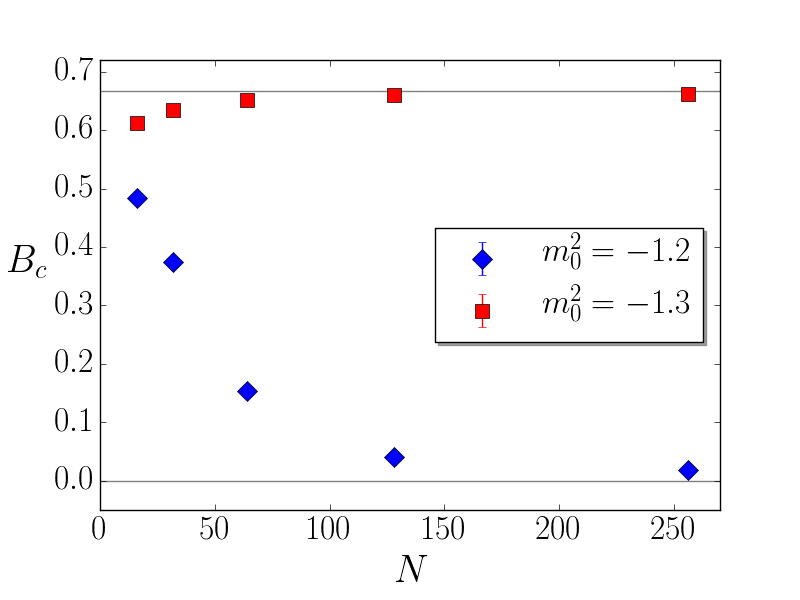}
\end{minipage}%
\caption{\label{fig:phases} Testing the two phases:
({\em Left panels}) Histograms representing the distribution of $\phi$ values 
on a single lattice in two regions 
corresponding to ({\em far left}) the symmetric phase ($am_0^2 = -0.64$, 
$\lambda = 
0.5$) 
and ({\em centre left}) the broken phase ($am_0^2 = -0.72$, $\lambda = 0.5$), 
in the 
infinite volume limit.
({\em Right panels}) The Binder cumulant, $B_c$, as a function of ({\em 
centre right}) 
the bare mass, $m_0^2$, 
at $\lambda=0.5$ and for different lattice volumes, and ({\em far right}) as a 
function 
of lattice size, at fixed bare mass and coupling constant. The horizontal grey 
lines correspond to the infinite volume values in the symmetric and broken 
phases: $B_c = 0$ and $B_c = 2/3$, respectively. The vertical red dashed line 
in the left-hand plot corresponds to the critical mass. Uncertainties are 
smaller 
than the size of the data point markers. Data from 10$^4$ measurements.
}
\end{figure}

I illustrate the existence of the phase transition, which occurs at $m_0^2 = 
-0.72$ for $\lambda = 0.5$ and at $m_0^2 = -1.27$ for $\lambda = 1.0$, in 
Figure \ref{fig:phases}. In the left-hand panels, I plot the distributions of 
the 
values of $\phi$ from a single lattice at 
two points in the mass-coupling parameter space. These points 
correspond to 
the symmetric ({\em far left}) and broken ({\em centre left}) phases in the 
infinite volume limit. The histograms demonstrate that, as 
the lattice size increases, the distribution 
of $\phi$ values collapses to a Gaussian distribution centred on zero for 
$m_0^2 = -0.64$, corresponding to the symmetric phase, and a Gaussian 
centred around $\langle \phi\rangle \neq 0$ for 
$m_0^2 = -0.72$, corresponding to the 
broken phase.

As a further test, I studied the fourth-order Binder cumulant, $B_c$, an order 
parameter for the 
phase 
transition from the symmetric to the broken phase.  This cumulant is defined 
as $B_c=1-\overline{\phi^4}/3(\overline{\phi^2})^2$, where 
$\overline{\phi}$ is the volume-averaged value of $\phi$ on a single 
configuration \cite{Binder:1981ul}.
In the infinite volume limit, $B_c = 0$ in the symmetric 
phase and $B_c = 2/3$ in the broken phase. In the right-hand panels of Figure 
\ref{fig:phases} I plot: ({\em centre right}) the Binder cumulant as a function 
of the bare lattice 
mass 
$m_0^2$, at fixed coupling constant $\lambda = 0.5$ and for different 
lattice 
sizes; and ({\em far right}) as a function of lattice size for three different 
bare lattice masses at fixed coupling $\lambda = 1.0$. These plots demonstrate 
that, in the infinite volume limit, the Binder cumulant tends to the correct 
value in both phases.

\paragraph{Mixing and the gradient flow}

Let us consider the matrix element $\left\langle \Omega | 
\phi^2(0) \cdot \phi(0) 
\partial^2 \phi(0) |\Omega \right\rangle$ in perturbation theory 
\cite{Monahan:2015lha}. In 
the continuum this matrix element vanishes. On 
the lattice, however, the corresponding matrix element in four 
dimensions diverges with the inverse lattice spacing squared, 
signaling the appearance of power-divergent 
mixing. In the continuum limit, 
keeping the flow time $\tau$ fixed in physical units, this matrix element 
tends to a constant, signaling the suppression of power-divergent mixing for 
smeared degrees of freedom. Note, however, that we have not removed mixing 
entirely, but only suppressed the problematic \emph{power-divergent} mixing. 
In two dimensions, the lattice matrix element diverges only 
logarithmically with the cutoff, but the principle remains the same: fixing the 
flow time removes this divergent behaviour.

\begin{figure}
\centering
\begin{minipage}{0.5\textwidth}
\includegraphics[width=0.95\textwidth,keepaspectratio=true]{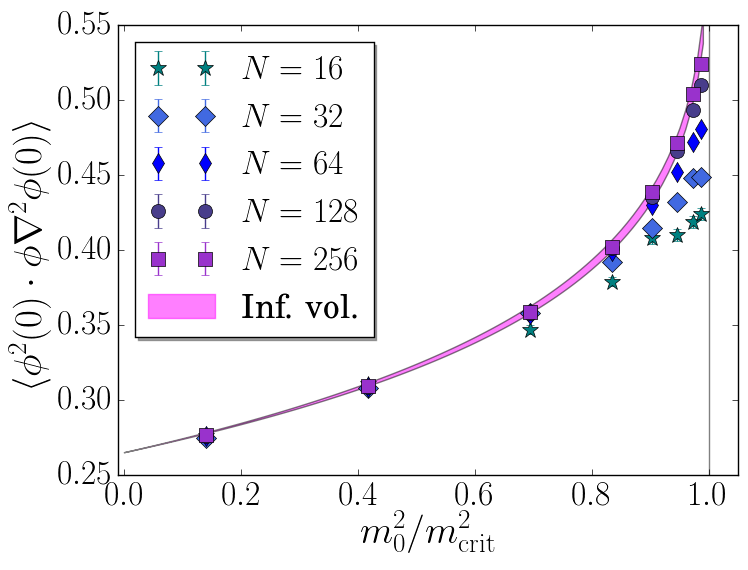}
\end{minipage}%
\begin{minipage}{0.5\textwidth}
\includegraphics[width=0.95\textwidth,keepaspectratio=true]{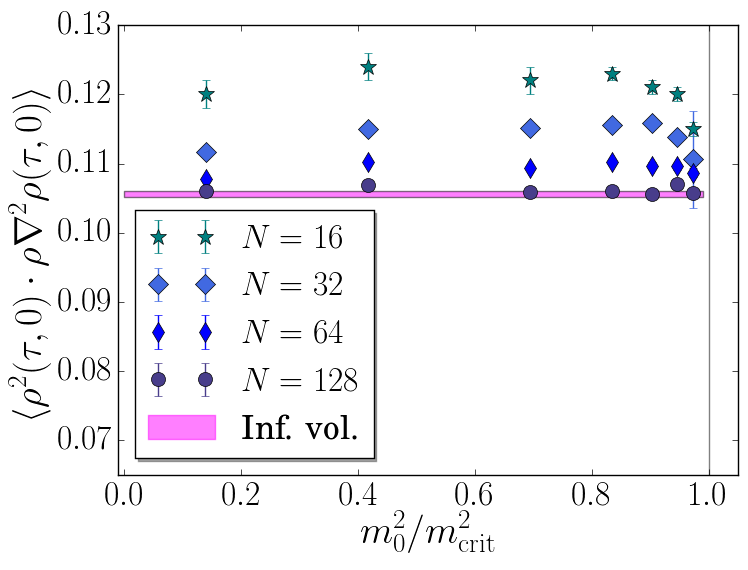}
\end{minipage}%
\caption{\label{fig:tw2} 
({\em Left}) The matrix element $\left\langle \Omega | 
\phi^2(0) \cdot \phi(0) 
\nabla^2 \phi(0) |\Omega \right\rangle$, as a function of the 
bare mass, expressed as a ratio of the critical mass, at fixed coupling. The 
magenta band shows a fit to a logarithmic function of 
$(1-m_0^2/m_{\mathrm{crit}}^2)$ for the infinite volume results, obtained from a 
polynomial fit in $1/N$. ({\em Right}) The same 
matrix element, but at non-zero flow time. Here the magenta band is a fit to a 
constant. Including terms polynomial in the flow time does not change the final 
fit, within errors.}
\end{figure}

We see this in Figure \ref{fig:tw2}. In the left-hand panel I plot the 
vacuum expectation value of $\phi^2(0) \cdot \phi(0) 
\nabla^2 \phi(0)$ in the symmetric phase, as a function of the bare mass $m_0^2$
(expressed as a ratio to the critical mass, $m_{\mathrm{crit}}^2$). The 
true continuum limit corresponds to the origin in the bare mass-coupling 
plane, but for our purposes we can consider instead the approach to the 
critical point at fixed coupling. I extrapolate to the infinite volume 
limit with a polynomial fit to $1/N$, including terms up to $1/N^3$, for 
lattices from $N=16$ to $N=256$. The magenta band corresponds to a fit to a 
logarithmic function of $(1-m_0^2/m_{\mathrm{crit}}^2)$ and incorporates fitting 
errors only. This plot shows the logarithmic divergence expected on 
perturbative grounds.

In the right-hand panel I plot the vacuum matrix element of $\rho^2(\tau,0) 
\cdot \rho(\tau,0) 
\nabla^2 \rho(\tau,0)$ as a function of the bare mass, at fixed flow time (for 
lattices up to $N=128$). I extrapolate to the infinite volume limit with a 
polynomial, as before, and fit these infinite volume results to a constant. 
Including terms linear or quadratic in the flow time has no affect on the 
central value of the fit, within errors.

These two plots demonstrate the power of the gradient flow: the smeared matrix 
elements no longer diverge in the continuum limit, exactly in line with 
perturbative expectations.

\section{Conclusion}
Power-divergent mixing complicates the continuum limit of lattice matrix 
elements for a wide range of calculations. The gradient flow provides a 
tool to remove power-divergent mixing from lattice calculations. Here I 
demonstrate this principle in the simple toy model of scalar field theory, with 
quartic interactions, in two dimensions. By keeping the 
flow time fixed in physical units in the continuum limit, power-divergent 
coefficients are rendered finite, at the expense of 
introducing a new physical scale into the system: the smearing radius.

One approach to 
dealing with this new scale is the smeared operator product expansion (sOPE), 
in which nonlocal operators are expanded in a basis of locally-smeared 
operators. The corresponding perturbative coefficients are then also functions 
of the smearing radius and, to a given order in perturbation theory and the 
flow time, the product of these perturbative coefficients with the matrix 
elements of smeared operators should be flow-time independent.

The sOPE also provides a natural framework for studying vacuum condensates and, 
in QCD, may provide sum-rule relations that relate vacuum matrix elements 
determined on the lattice to hadronic parameters. To extract meaningful 
physics, however, one must move beyond the toy model of two-dimensional scalar 
field theory, studied here, and implement the sOPE in QCD. This work is in 
progress.

\begin{acknowledgments}
I would like to thank Carleton DeTar, Herbert Neuberger and Kostas Orginos for 
many enlightening discussions during the course of this work, and Rich 
Brower and Mattia Dalla 
Brida for conversations regarding related calculations. 
This project was supported in part by the
U.S.~Department of Energy, Grant No.~DE-FG02-04ER41302 and in part by the 
U.S.~National Science Foundation under Grant No.~NSF PHY10-034278.
\end{acknowledgments}


\end{document}